\documentclass[12pt]{article}

\usepackage[vcentermath]{youngtab}

\author{Yu.~M.~Zinoviev
       \thanks{E-mail address: Yurii.Zinoviev@ihep.ru} \\
        {\it Institute for High Energy Physics} \\
        {\it of National Research Center "Kurchatov Institute"} \\
        {\it Protvino, Moscow Region, 142280, Russia}}
\title{Infinite spin fields in $d=3$ and beyond}

\date{}

\begin{document}

\maketitle

\begin{abstract}
In this paper we consider the frame-like formulation for the
so called infinite (continuous) spin representations of the Poincare
algebra. In the tree dimensional case we give explicit Lagrangian
formulation for bosonic and fermionic infinite spin fields (including
the complete sets of the gauge invariant objects and all the necessary
extra fields). Moreover we find the supertransformations for the
supermultiplet containing one bosonic and one fermionic fields leaving
the sum of their Lagrangians invariant. Properties of such fields and
supermultiplets in four and higher dimensions are also briefly
discussed.
\end{abstract}

\thispagestyle{empty}
\newpage
\setcounter{page}{1}

\section*{Introduction}

Besides the very well known finite components massless and massive
representations of the Poincare algebra there are rather exotic so
called infinite (or continuous) spin ones (see e.g.
\cite{BKRX02,BB06}). In dimensions $d \ge 4$ they have infinite number
of physical degrees of freedom and so may be of some interest for the
higher spins theory. Indeed they attracted some attention last times
\cite{BM05,Ben13,ST14,ST14a,Riv14,BNS15}. It has been noted several
times that such infinite spin representations may be considered as a
limit of massive higher spin ones where spin goes to infinity, mass
goes to zero while they product being fixed. Moreover, recently
Metsaev has shown the the metric-like Lagrangian formulation for the
bosonic \cite{Met16} and fermionic \cite{Met17} fields in $AdS_d$
spaces with $d \ge 4$ can be constructed using exactly the same
technique as was previously used for the gauge invariant formulation
of massive higher spin bosonic \cite{Zin01} and fermionic
\cite{met06} fields. 

The current paper is devoted to the frame-like formulation for such
infinite spin fields. In the first (and main) section we construct
gauge invariant Lagrangian formulation for bosonic and fermionic
cases. We also elaborate on the whole set of the gauge invariant
objects (introducing all necessary extra fields) and rewrite our
Lagrangians in  the explicitly gauge invariant form. Moreover we
managed to find supertransformations for the supermultiplet containing
one bosonic and one fermionic infinite spin fields that leaves the sum
of their Lagrangians invariant. For this we heavily use our previous
results on the gauge invariant formulation for massive bosonic and
fermionic higher spin fields in $d=3$ \cite{BSZ12a,BSZ14a} (see also
\cite{BPSS15,Zin15,Zin16}) as well as on the massive higher spin
supermultiplets \cite{BSZ15,BSZ16,BSZ17}. In the last two sections we
briefly discuss the properties of such fields and supermultiplets in
$d=4$ and $d \ge 5$ dimensions leaving explicit details to the
forthcoming publication.

\noindent
{\bf Notations and conventions} We will work in the frame-like
multispinor formalism where all objects are two, one or zero forms
with a set of completely symmetric local spinor indices. We will
follow mostly the conventions of \cite{BSZ17} but we restrict
ourselves with the flat Minkowski space.

\section{Infinite spin fields in $d=3$}

In this section we develop the frame like formalism for the massless
infinite spin bosonic and fermionic fields as well as for the
supermultiplet containing such fields.

\subsection{Infinite spin boson}

As we have already noted there is a tight connection between the gauge
invariant description for the massive finite spin fields and the one
for the massless infinite spin ones. Thus we will follow the same
approach as in \cite{BSZ12a} but this time without restriction on the
number of components. So we introduce an infinite set of physical and
auxiliary one-forms $\Omega^{\alpha(2k)}$, $\Phi^{\alpha(2k)}$,
$1 \le k \le \infty$ as well as one-form $A$ and zero-forms
$B^{\alpha(2)}$, $\pi^{\alpha(2)}$ and $\varphi$\footnote{Note that in
three dimensions such infinite spin bosonic field (as any massive
higher spin boson) has just two physical degrees of freedom, while
infinite spin fermionic field (as any massive higher spin fermion) has
just one. However it is impossible to realize such representations
using a finite number of components (see e.g. \cite{ST14a}.}. We begin
with the sum of kinetic terms for all these fields:
\begin{eqnarray}
{\cal L}_0 &=& \sum_{k=1}^\infty (-1)^{k+1} [ k 
\Omega_{\alpha(2k-1)\beta} e^\beta{}_\gamma 
\Omega^{\alpha(2k-1)\gamma} + \Omega_{\alpha(2k)} d \Phi^{\alpha(2k)}
] \nonumber \\
 && + E B_{\alpha(2)} B^{\alpha(2)} - B_{\alpha(2)} e^{\alpha(2)} d A
- E \pi_{\alpha(2)} \pi^{\alpha(2)} + \pi_{\alpha(2)} E^{\alpha(2)} d
\varphi
\end{eqnarray}
as well as their initial gauge transformations:
\begin{equation}
\delta_0 \Omega^{\alpha(2k)} = d \eta^{\alpha(2k)}, \qquad
\delta_0 \Phi^{\alpha(2k)} = d \xi^{\alpha(2k)} + e^\alpha{}_\beta
\eta^{\alpha(2k-1)\beta}, \qquad
\delta_0 A = d \xi
\end{equation}
Then following general scheme we add to the Lagrangian a set of cross
terms gluing all these components together:
\begin{eqnarray}
{\cal L}_1 &=& \sum_{k=1}^\infty (-1)^{k+1} [ \tilde{a}_k
\Omega_{\alpha(2k)\beta(2)} e^{\beta(2)} \Phi^{\alpha(2k)} + a_k
\Omega_{\alpha(2k)} e_{\beta(2)} \Phi^{\alpha(2k)\beta(2)} ] \nonumber
\\
 && + \tilde{a}_0 \Omega_{\alpha(2)} e^{\alpha(2)} A - a_0
\Phi_{\alpha\beta} E^\beta{}_\gamma B^{\alpha\gamma} + \hat{a}_0
\pi_{\alpha(2)} E^{\alpha(2)} A
\end{eqnarray}
and introduce appropriate corrections for the gauge transformations:
\begin{eqnarray}
\delta_1 \Omega^{\alpha(2k)} &=& \frac{(k+2)}{k}a_k e_{\beta(2)}
\eta^{\alpha(2k)\beta(2)} + \frac{a_{k-1}}{k(2k-1)} e^{\alpha(2)}
\eta^{\alpha(2k-2)} \nonumber \\
\delta_1 \Phi^{\alpha(2k)} &=& a_k e_{\beta(2)} 
\xi^{\alpha(2k)\beta(2)} + \frac{(k+1)a_{k-1}}{k(k-1)(2k-1)}
e^{\alpha(2)} \xi^{\alpha(2k-2)} \nonumber \\
\delta_1 \Omega^{\alpha(2)} &=& 3a_1 e_{\beta(2)} 
\eta^{\alpha(2)\beta(2)}, \qquad
\delta_1 \Phi^{\alpha(2)} = a_1 e_{\beta(2)} \xi^{\alpha(2)\beta(2)}
+ 2a_0 e^{\alpha(2)} \xi \\
\delta_1 B^{\alpha(2)} &=& 2a_0 \eta^{\alpha(2)}, \qquad
\delta_1 A = \frac{a_0}{4} e_{\alpha(2)} \xi^{\alpha(2)}, \qquad
\delta_1 \varphi = - \hat{a}_0 \xi \nonumber
\end{eqnarray}
Here consistency of the gauge transformations with the Lagrangian
requires:
$$
\tilde{a}_k = - \frac{(k+2)}{k} a_k, \qquad \tilde{a}_0 = 2a_0
$$
At last we introduce mass-like terms for all components and
appropriate corrections to the gauge transformations:
\begin{equation}
{\cal L}_2 = \sum_{k=1}^\infty (-1)^{k+1} b_k \Phi_{\alpha(2k-1)\beta}
e^\beta{}_\gamma \Phi^{\alpha(2k-1)\gamma} + b_0 \Phi_{\alpha(2)}
E^{\alpha(2)} \varphi + \tilde{b}_0 E \varphi^2
\end{equation}
\begin{equation}
\delta_2 \Omega^{\alpha(2k)} = \frac{b_k}{k} e^\alpha{}_\beta
\xi^{\alpha(2k-1)\beta}, \qquad
\delta_2 \pi^{\alpha(2)} = b_0 \xi^{\alpha(2)}
\end{equation}
Now we require that the whole Lagrangian ${\cal L} = {\cal L}_0 +
{\cal L}_1 + {\cal L}_2$ will be invariant under the gauge
transformations $\delta = \delta_0 + \delta_1 + \delta_2$. This
produce the following general relations on the parameters:
\begin{equation}
(k+2)^2 b_{k+1} = k(k+1)b_k
\end{equation}
\begin{equation}
\frac{2(k+2)(2k+3)}{(k+1)(2k+1)} a_k{}^2 - \frac{2(k+1)}{(k-1)} 
a_{k-1}{}^2 + 4b_k = 0
\end{equation}
as well as some relations for the lower components: 
$$
5a_1{}^2 - a_0{}^2 + 4b_1 = 0
$$
$$
\hat{a}_0{}^2 = 64b_1, \qquad
b_0 = \frac{\hat{a}_0a_0}{4}, \qquad
\tilde{b}_0 = \frac{3a_0{}^2}{2}
$$
The general solution of all these relations has two free parameters.
In the massive finite spin case its just the mass and spin but in our
case we choose $a_0$ and $b_1$ as the main ones. Then all other
parameters can be expressed as follows:
\begin{equation}
b_k = \frac{4b_1}{k(k+1)^2}
\end{equation}
\begin{equation}
a_k{}^2 = \frac{k}{(2k+3)} [ \frac{3(k+1)}{2(k+2)}a_0{}^2 - 
\frac{8k}{(k+1)} b_1 ]
\end{equation}

Now we are ready to analyze the solution obtained. Let us begin with
the case $a_0{}^2 < 16b_1$. In general it means that starting from
some value of $k$ all $a_k{}^2$ become negative so that we obtain non
unitary theory. The only exceptions happen then one adjust the values
of $a_{}^2$ and $b_1$ so that at some $k_0$ we obtain $c_{k_0} = 0$.
In this case we obtain unitary theory with the finite number of
components and this case corresponds to the gauge invariant
description for the massive bosonic field with the spin $k_0+1$. 
Let us turn to the case $a_0{}^2 = 16b_1$ (this corresponds to the
case $\mu_0=0$ in \cite{Met16}). In this case we obtain:
\begin{equation}
a_k{}^2 = \frac{3k}{2(k+1)(k+2)(2k+3)} a_0{}^2
\end{equation}
so we get an unitary theory with infinite number of components. Note
that for the case $a_0{}^2 > 16b_1$ we also obtain unitary theory but
as it was shown by Metsaev \cite{Met16} it corresponds to the
tachionic infinite spin field. Thus in what follows we will restrict
ourselves with the case $a_0{}^2 = 16b_1$ only.

One of the nice and general features of the frame like formalism is
that for each field (physical or auxiliary) one can construct
corresponding gauge invariant object. For the case at hands we will
follow massive case in \cite{Zin16,BSZ17}. For almost all fields
corresponding gauge invariant objects can be directly constructed from
the known form for the gauge transformations given above (here for the
later convenience we chaged normalization for the zero-forms
$B^{\alpha(2)} \Rightarrow 2a_0 B^{\alpha(2)}$,
$\pi^{\alpha(2)} \Rightarrow b_0 \pi^{\alpha(2)}$):
\begin{eqnarray}
{\cal R}^{\alpha(2k)} &=& d \Omega^{\alpha(2k)} + \frac{b_k}{k} 
e^\alpha{}_\beta \Phi^{\alpha(2k-1)\beta} 
+ \frac{(k+2)}{k}a_k e_{\beta(2)} \Omega^{\alpha(2k)\beta(2)}
+ \frac{a_{k-1}}{k(2k-1)} e^{\alpha(2)} \Omega^{\alpha(2k-2)}
\nonumber \\
{\cal T}^{\alpha(2k)} &=& d \Phi^{\alpha(2k)} + e^\alpha{}_\beta
\Omega^{\alpha(2k-1)\beta} + a_k e_{\beta(2)} 
\Phi^{\alpha(2k)\beta(2)} +
\frac{(k+1)a_{k-1}}{k(k-1)(2k-1)}e^{\alpha(2)} \Phi^{\alpha(2k-2)}
\nonumber \\
{\cal R}^{\alpha(2)} &=& d \Omega^{\alpha(2)} + b_1 e^\alpha{}_\beta
\Phi^{\alpha\beta} + 3a_1 e_{\beta(2)} \Omega^{\alpha(2)\beta(2)} 
- a_0{}^2 E^\alpha{}_\beta B^{\alpha\beta} + b_0 E^{\alpha(2)} \varphi
\nonumber \\
{\cal T}^{\alpha(2)} &=& d \Phi^{\alpha(2)} + e^\alpha{}_\beta
\Omega^{\alpha\beta} + a_1 e_{\beta(2)} \Phi^{\alpha(2)\beta(2)} +
2a_0 e^{\alpha(2)} A \\
{\cal A} &=& d A - 2a_0 E_{\alpha(2)} B^{\alpha(2)} + \frac{a_0}{4}
e_{\alpha(2)} \Phi^{\alpha(2)} \nonumber  \\
\Phi &=& d \varphi - \frac{\sqrt{3}}{2}a_0{}^2 e_{\alpha(2)}
\pi^{\alpha(2)} + 2\sqrt{3}a_0 A  \nonumber
\end{eqnarray}
But to construct gauge invariant objects for $B^{\alpha(2)}$ and 
$\pi^{\alpha(2)}$ one must introduce a first pair of the so called
extra fields\footnote{Recall that extra fields are the fields that do
not enter the free Lagrangian but are necessary for the construction
of the whole set of gauge invariant objects. Moreover such fields play
important role in the construction of the interactions.}
$B^{\alpha(4)}$ and $\pi^{\alpha(4)}$:
\begin{eqnarray}
{\cal B}^{\alpha(2)} &=& d B^{\alpha(2)} - \Omega^{\alpha(2)} 
+ b_1 e^\alpha{}_\beta \pi^{\alpha\beta} + 3a_1 e_{\beta(2)}
B^{\alpha(2)\beta(2)} \nonumber \\
\Pi^{\alpha(2)} &=& d \pi^{\alpha(2)} + e^\alpha{}_\beta 
B^{\alpha\beta} - \Phi^{\alpha(2)} - \frac{1}{\sqrt{3}} e^{\alpha(2)}
\varphi + a_1 e_{\beta(2)} \pi^{\alpha(2)\beta(2)} 
\end{eqnarray}
which transform as follows:
$$
\delta B^{\alpha(4)} = \eta^{\alpha(4)}, \qquad
\delta \pi^{\alpha(4)} = \xi^{\alpha(4)}
$$
But to construct gauge invariant objects for these new fields one must
introduce the next pair of extra fields and so on. This results in the
infinite chain of zero forms $B^{\alpha(2k)}$ and $\pi^{\alpha(2k)}$,
$1 \le k \le \infty$ with the following set of gauge invariant
objects:
\begin{eqnarray}
{\cal B}^{\alpha(2k)} &=& d B^{\alpha(2k)} - \Omega^{\alpha(2k)}
+ \frac{b_k}{k} e^\alpha{}_\beta \pi^{\alpha(2k-1)\beta} +
\frac{(k+2)}{k}a_k e_{\beta(2)} B^{\alpha(2k)\beta(2)} \nonumber \\
 && +  \frac{a_{k-1}}{k(2k-1)} e^{\alpha(2)} B^{\alpha(2k-2)}
\nonumber \\
\Pi^{\alpha(2k)} &=& d \pi^{\alpha(2k)} - \Phi^{\alpha(2k)} + 
e^\alpha{}_\beta B^{\alpha(2k-1)\beta} + a_k e_{\beta(2)}
\pi^{\alpha(2k)\beta(2)} \\
 && + \frac{(k+1)a_{k-1}}{k(k-1)(2k-1)} e^{\alpha(2)}
\pi^{\alpha(2k-2)} \nonumber
\end{eqnarray}
Here:
$$
\delta B^{\alpha(2k)} = \eta^{\alpha(2k)}, \qquad
\delta \pi^{\alpha(2k)} = \xi^{\alpha(2k)}
$$

Now we have an infinite set of gauge one-forms as well as an infinite
set of Stueckelberg zero-forms. As in the massive finite spin case
\cite{Zin16,BSZ17} this allows us to rewrite the Lagrangian in the
explicitly gauge invariant form:
\begin{equation}
{\cal L} = - \frac{1}{2} \sum_{k=1}^\infty (-1)^{k+1} [ 
{\cal R}_{\alpha(2k)} \Pi^{\alpha(2k)} + {\cal T}_{\alpha(2k)}
{\cal B}^{\alpha(2k)}] + \frac{1}{2} e_{\alpha(2)}
{\cal B}^{\alpha(2)} \Phi
\end{equation}
By construction each term here is separately gauge invariant and the
explicit values for all coefficients are determined by the so called
extra field decoupling conditions:
$$
\frac{\delta {\cal L}}{\delta B^{\alpha(2k)}} = 0, \qquad
\frac{\delta {\cal L}}{\delta \pi^{\alpha(2k)}} = 0, \qquad
2 \le k \le \infty
$$

\subsection{Fermionic case}

In this case we will also follow the construction for the massive
finite spin field \cite{BSZ14a} but this time for the infinite set of
components. So we introduce a set of one-forms $\Psi^{\alpha(2k+1)}$,
$0 \le k \le \infty$ and a zero-form $\psi^\alpha$. Once again we
begin with the sum of kinetic terms for all fields:
\begin{equation}
\frac{1}{i}{\cal L}_0 = \sum_{k=0}^\infty \frac{(-1)^{k+1}}{2}
\Psi_{\alpha(2k+1)} d \Psi^{\alpha(2k+1)} + \frac{1}{2} \psi_\alpha 
E^\alpha{}_\beta d \psi^\beta
\end{equation}
as well as with their initial gauge transformations:
\begin{equation}
\delta_0 \Psi^{\alpha(2k+1)} = d \zeta^{\alpha(2k+1)}
\end{equation}
Now we add a set of cross terms gluing them together
\begin{equation}
\frac{1}{i}{\cal L}_1 = \sum_{k=1}^\infty (-1)^{k+1} c_k
\Psi_{\alpha(2k-1)\beta(2)} e^{\beta(2)} \Psi^{\alpha(2k-1)} + c_0
\Psi_\alpha E^\alpha{}_\beta \psi^\beta
\end{equation}
and corresponding corrections to the gauge transformations:
\begin{eqnarray}
\delta_1 \Psi^{\alpha(2k+1)} &=& c_{k+1} e_{\beta(2)} 
\zeta^{\alpha(2k+1)\beta(2)} + \frac{c_k}{k(2k+1)} e^{\alpha(2)}
\zeta^{\alpha(2k-1)}, \nonumber \\
\delta_1 \psi^\alpha &=& c_0 \zeta^\alpha
\end{eqnarray}
At last we add the mass-like terms for all fields and appropriate
corrections to the gauge transformations:
\begin{equation}
\frac{1}{i}{\cal L}_2 = \sum_{k=0}^\infty (-1)^{k+1} \frac{d_k}{2}
\Psi_{\alpha(2k)\beta} e^\beta{}_\gamma \Psi^{\alpha(2k)\gamma} -
\frac{m_0}{2} E \psi_\alpha \psi^\alpha 
\end{equation}
\begin{equation}
\delta_2 \Psi^{\alpha(2k+1)} = \frac{d_k}{(2k+1)} e^\alpha{}_\beta
\zeta^{\alpha(2k)\beta}
\end{equation}
Now we require that the whole Lagrangian ${\cal L} = {\cal L}_0 +
{\cal L}_1 + {\cal L}_2$ will be invariant under the gauge
transformations $\delta = \delta_0 + \delta_1 + \delta_2$. This
produce a number of general relations on the parameters
\begin{equation}
(2k+5)d_{k+1} = (2k+3) d_k
\end{equation}
\begin{equation}
\frac{(k+2)(2k+1)}{(k+1)(2k+3)} c_{k+1}{}^2 - c_k{}^2 + 
\frac{d_k{}^2}{(2k+1)} = 0
\end{equation}
as well as
$$
\frac{8}{3}c_1{}^2 - c_0{}^2 + 4d_0{}^2 = 0, \qquad
d_0 = \frac{m_0}{3}
$$
As in the bosonic case the general solution for all these relations
has two free parameters and we choose $c_0$ and $m_0$ this time. Then
all other coefficients can be expressed as follows:
\begin{equation}
d_k = \frac{m_0}{(2k+3)}
\end{equation}
\begin{equation}
c_k{}^2 = \frac{(2k+1)}{4(k+1)} c_0{}^2 - \frac{k}{2(2k+1)} m_0{}^2
\end{equation}

The properties of this solution appears to be the same as in the
bosonic case. Namely, for the case $m_0{}^2 > 2c_0{}^2$ in general we
obtain non unitary theory. The only exceptions appear if one adjust
this parameters so that at some $k_0$ we get $c_{k_0} = 0$. In this
case we obtain unitary theory with finite number of components which
corresponds to the gauge invariant description for massive fermionic
field with spin $k_0 + 3/2$. For the $m_0{}^2 = 2c_0{}^2$ (this
corresponds to $\mu_0 = 0$ in \cite{Met17}) we obtain
\begin{equation}
c_k{}^2 = \frac{c_0{}^2}{4(k+1)(2k+1)}
\end{equation}
that corresponds to the unitary massless infinite spin field while for
the $m_0{}^2 < c_0{}^2$ we again obtain tachionic infinite spin case.
As in the bosonic case in what follows we will restrict ourselves with
the case $m_0{}^2 = 2c_0{}^2$ only.

Now we proceed with the construction of the full set of gauge
invariant objects. For all one-forms the construction is pretty
straightforward (again for the later convenience we changed
normalization for the zero-form $\psi^\alpha \Rightarrow c_0 
\psi^\alpha$):
\begin{eqnarray}
{\cal F}^{\alpha(2k+1)} &=& d \Psi^{\alpha(2k+1)} + \frac{d_k}{(2k+1)}
e^\alpha{}_\beta \Psi^{\alpha(2k)\beta} + c_{k+1} e_{\beta(2)}
\Psi^{\alpha(2k+1)\beta(2)} \nonumber \\
 && + \frac{c_k}{k(2k+1)} e^{\alpha(2)} \Psi^{\alpha(2k-1)} \\
{\cal F}^\alpha &=& D \Psi^\alpha + d_0 e^\alpha{}_\beta \Psi^\beta +
c_1 e_{\beta(2)} \Psi^{\alpha\beta(2)} - c_0{}^2 E^\alpha{}_\beta 
\psi^\beta \nonumber
\end{eqnarray}
But to construct gauge invariant object for the zero-form one must
introduce a first extra field:
\begin{equation}
{\cal C}^\alpha = d \psi^\alpha - \Psi^\alpha + d_0 
e^\alpha{}_\beta \psi^\beta + c_1 e_{\beta(2)} \psi^{\alpha\beta(2)},
\qquad \delta \psi^{\alpha(3)} = \zeta^{\alpha(3)}
\end{equation}
Then to construct gauge invariant object for this field one must
introduce the second one and so on. This results in the infinite set
of zero-forms with the corresponding gauge invariant objects:
\begin{eqnarray}
{\cal C}^{\alpha(2k+1)} &=& d \psi^{\alpha(2k+1)} - 
\Psi^{\alpha(2k+1)} + \frac{d_k}{(2k+1)} e^\alpha{}_\beta
\psi^{\alpha(2k)\beta} + c_{k+1} e_{\beta(2)}
\psi^{\alpha(2k+1)\beta(2)} \nonumber \\
 && + \frac{c_k}{k(2k+1)} e^{\alpha(2)}
\psi^{\alpha(2k-1)}  
\end{eqnarray}
where
$$
\delta \psi^{\alpha(2k+1)} = \zeta^{\alpha(2k+1)}
$$

Now we have an infinite set of one-form and zero-form fields and their
gauge invariant two and one forms. This allows us to rewrite the
Lagrangian in the explicitly gauge invariant form:
\begin{equation}
{\cal L} = - \frac{i}{2} \sum_{k=0}^\infty (-1)^{k+1} 
{\cal F}_{\alpha(2k+1)} {\cal C}^{\alpha(2k+1)}
\end{equation}
As in the bosonic case each term is separately gauge invariant while
the specific values of all coefficients are determined by the extra
field decoupling condition:
$$
\frac{\delta{\cal L}}{\delta \psi^{\alpha(2k+1)}} = 0, \qquad
1 \le k \le \infty
$$

\subsection{Infinite spin supermultiplet}

It is interesting (see e.g. \cite{BKRX02}) that similarly to the usual
massless and massive fields such massless infinite spin fields also
can form supermultiplets. In $d=3$ the minimal supermultiplets
contains just one bosonic and one fermionic fields. Due to tight
relation with gauge invariant formulation for the massive higher spin
fields and supermultiplets here we will heavily use the results of our
recent paper \cite{BSZ17}. The main difference (besides the infinite
set of components) is the essentially different expressions for the
coefficients $a_k$ and $c_k$.

The general strategy will be to find explicit form of the
supertransformations for all fields such that all gauge invariant two
and one forms transform covariantly and to check the invariance of the
Lagrangian. Let us begin with the bosonic fields. For the general case
$k \ge 2$ we will use the following ansatz:
\begin{eqnarray}
\delta \Omega^{\alpha(2k)} &=& i\rho_k \Psi^{\alpha(2k-1)} 
\zeta^\alpha + i\sigma_k \Psi^{\alpha(2k)\beta} \zeta_\beta \nonumber
\\
\delta \Phi^{\alpha(2k)} &=& i\alpha_k \Psi^{\alpha(2k-1)} 
\zeta^\alpha + i\beta_k \Psi^{\alpha(2k)\beta} \zeta_\beta
\end{eqnarray}
and require that the corresponding two-form transform covariantly:
\begin{eqnarray}
\delta {\cal R}^{\alpha(2k)} &=& i\rho_k {\cal F}^{\alpha(2k-1)} 
\zeta^\alpha + i\sigma_k {\cal F}^{\alpha(2k)\beta} \zeta_\beta
\nonumber \\
\delta {\cal T}^{\alpha(2k)} &=& i\alpha_k {\cal F}^{\alpha(2k-1)} 
\zeta^\alpha + i\beta_k {\cal F}^{\alpha(2k)\beta} \zeta_\beta
\end{eqnarray}
First of all this gives us an important relation
\begin{equation}
c_0{}^2 = 6a_0{}^2
\end{equation}
Recall that the parameters $a_0$ and $c_0$ are the main dimension-full
parameters that determine the whole construction for the bosonic and
fermionic fields. So this relation plays the same role as the
requirement that masses of bosonic and fermionic fields in the
supermultiplet must be equal. Further, we obtain explicit expressions
for all parameters
$$
\alpha_k{}^2 = k \hat{\alpha}^2, \qquad
\beta_k{}^2 = \frac{(k+1)}{2k(2k+1)} \hat{\alpha}^2
$$
$$
\sigma_k{}^2 = \frac{3a_0{}^2}{4k(k+1)^2} \hat{\alpha}^2, \qquad
\rho_k{}^2 = \frac{3a_0{}^2}{8k^3(k+1)(2k+1)} \hat{\alpha}^2
$$
where $\hat{\alpha}$ is an arbitrary parameter that can be fixed by
the normalization of the superalgebra. 

For the three bosonic components that require separate consideration
we obtain:
\begin{eqnarray}
\delta \Omega^{\alpha(2)} &=& i\rho_1 \Psi^\alpha \zeta^\alpha +
i\sigma_1 \Psi^{\alpha(2)\beta} \zeta_\beta - 
\frac{i\sqrt{3}a_0{}^2}{4}\hat{\alpha} e^{\alpha(2)} \psi^\beta 
\zeta_\beta \nonumber  \\
\delta A &=& \frac{i\hat{\alpha}}{2} \Psi^\alpha \zeta_\alpha +
\frac{i\sqrt{3}a_0}{2}\hat{\alpha} \psi_\alpha e^{\alpha\beta} 
\zeta_\beta, \qquad
\delta \varphi = - i\sqrt{3}a_0\hat{\alpha} \psi^\alpha \zeta_\alpha
\end{eqnarray}
At last the supertransformations for the zero-forms look like:
\begin{eqnarray}
\delta B^{\alpha(2k)} &=& i\sigma_k \psi^{\alpha(2k)\beta} \zeta_\beta
+ i\rho_k \psi^{\alpha(2k-1)} \zeta^\alpha \nonumber \\
\delta \pi^{\alpha(2k)} &=& i\beta_k \psi^{\alpha(2k)\beta} 
\zeta_\beta + i\alpha_k \psi^{\alpha(2k-1)} \zeta^\alpha
\end{eqnarray}
where all coefficients $\alpha_k$, $\beta_k$, $\rho_k$ and $\sigma_k$
are the same as above.

Now let us turn to the fermionic components. For the general case 
$k \ge 1$ we will consider the following ansatz:
\begin{eqnarray}
\delta \Psi^{\alpha(2k+1)} &=& \frac{\alpha_k}{(2k+1)}
\Omega^{\alpha(2k)} \zeta^\alpha + 2(k+1)\beta_{k+1} 
\Omega^{\alpha(2k+1)\beta} \zeta_\beta \nonumber \\
 && + \gamma_k \Phi^{\alpha(2k)} \zeta^\alpha + \delta_k
\Phi^{\alpha(2k+1)\beta} \zeta_\beta
\end{eqnarray}
Then the requirement that the corresponding two-forms transform
covariantly:
\begin{eqnarray}
\delta {\cal F}^{\alpha(2k+1)} &=& \frac{\alpha_k}{(2k+1)}
{\cal R}^{\alpha(2k)} \zeta^\alpha + 2(k+1)\beta_{k+1} 
{\cal R}^{\alpha(2k+1)\beta} \zeta_\beta \nonumber \\
 && + \gamma_k {\cal T}^{\alpha(2k)} \zeta^\alpha + \delta_k
{\cal T}^{\alpha(2k+1)\beta} \zeta_\beta
\end{eqnarray}
gives us the same relation on the parameters $a_0$ and $c_0$ as before
and also gives:
$$
\gamma_k{}^2 = \frac{3a_0{}^2}{4k(k+1)^2(2k+1)^2} \hat{\alpha}^2
$$
$$
\delta_k{}^2 = \frac{3a_0{}^2}{2(k+1)(k+2)(2k+3)} \hat{\alpha}^2
$$
Again there is a couple of components that need to be considered
separately:
\begin{eqnarray}
\delta \Psi^\alpha &=& 2\beta_1 \Omega^{\alpha\beta} \zeta_\beta +
\delta_0 \Phi^{\alpha\beta} \zeta_\beta + a_0\hat{\alpha} e_{\beta(2)}
B^{\beta(2)} \zeta^\alpha + \sqrt{3}a_0\hat{\alpha} A \zeta^\alpha
- \frac{\sqrt{3}a_0}{2}\hat{\alpha} \varphi e^\alpha{}_\beta 
\zeta^\beta \nonumber \\
\delta \psi^\alpha &=& \frac{2\sqrt{3}}{3}\hat{\alpha} B^{\alpha\beta}
\zeta_\beta + \frac{a_0}{2}\hat{\alpha} \pi^{\alpha\beta} \zeta_\beta
+ \frac{\hat{\alpha}}{2} \varphi \zeta^\alpha
\end{eqnarray}
At last for the Stueckelberg zero-forms we obtain:
\begin{eqnarray}
\delta \psi^{\alpha(2k+1)} &=& \frac{\alpha_k}{(2k+1)} B^{\alpha(2k)}
\zeta^\alpha + 2(k+1)\beta_{k+1} B^{\alpha(2k+1)\beta} \zeta_\beta
\nonumber \\
 && + \gamma_k \pi^{\alpha(2k)} \zeta^\alpha + \delta_k 
\pi^{\alpha(2k+1)\beta} \zeta_\beta
\end{eqnarray}
where all parameters $\alpha_k$, $\beta_k$, $\gamma_k$ and $\delta_k$
are the same as before.

We have explicitly checked that the sum of the bosonic and fermionic
Lagrangians is invariant under these supertransformations up to the
terms proportional to the auxiliary fields $B^{\alpha(2)}$ and
$\pi^{\alpha(2)}$ equations in the same way as in the case of massive
higher spin supermultiplets \cite{BSZ17}.

\section{Infinite spin fields in $d=4$}

Similarly to the three dimensional case in $d=4$ there exist just one
bosonic and one fermionic infinite spin representations corresponding
to the completely symmetric (spin-)tensors. Metric-like gauge
invariant Lagrangian formulation (valid also in $d > 4$) has been
constructed recently \cite{Met16,Met17}. Frame-like Lagrangian
formulation can be straightforwardly obtained from the frame-like
gauge invariant formalism for the massive completely symmetric
(spin-)tensors developed in \cite{Zin08b}. These results will be
presented elsewhere.

The complete set of the gauge invariant objects for the massive
bosonic higher spin fields in $d \ge 4$ has been constructed in
\cite{PV10}. It requires the following three sets of fields:
$$
\Phi_\mu{}^{a(k),b(l)}, \quad S^{a(k),b(l)} \quad 0 \le k \le s-1,
\quad 0 \le l \le k
$$
$$
W^{a(k),b(l)} \quad k \ge s, \quad 0 \le l \le s-1
$$
where notation $\Phi_\mu{}^{a(k),b(l)}$ means that local indices
correspond to the Young tableau with two rows. Thus we have two finite
sets of gauge one-forms and Stueckelberg zero-forms as well as
infinite number of gauge invariant zero-forms. As in  the three
dimensional case one can try to consider the limit where spin goes to
infinity and mass goes to zero, but in $d > 3$ it appears to be rather
involved task. As for the analogous formulation for the massive
fermionic higher spin fields to the best of our knowledge it still
remain to be elaborated.

As it quite well known in $d=4$ there exist two type of massive higher
spin $N=1$ supermultiplets  corresponding to the integer or
half-integer superspins:
$$
\left( \begin{array}{c} s+\frac{1}{2} \\ s \qquad s' \\
 s-\frac{1}{2} \end{array} \right) \qquad
\left( \begin{array}{c} s+1 \\ s+\frac{1}{2} \qquad s+\frac{1}{2} 
\\  s \end{array} \right)
$$
Their explicit Lagrangian description was constructed in
\cite{Zin07a} using gauge invariant description for massive bosonic
and fermionic higher spin fields. The main idea was that massive
supermultiplet can be constructed out of the appropriately chosen set
of the massless ones. The decomposition of these two massive
supermultiplets into the massless one looks as follows:
$$
\left( \begin{array}{c} \Phi_{s+\frac{1}{2}} \\ A_s \qquad B_s \\
\Psi_{s-\frac{1}{2}} \end{array} \right) \qquad \Rightarrow \qquad
\sum_{k=1}^s \quad \left( \begin{array}{c} \Phi_{k+\frac{1}{2}} \\ A_k
\qquad  B_k \\ \Psi_{k-\frac{1}{2}} \end{array} \right) \quad \oplus
\quad \left( \begin{array}{c} \Phi_\frac{1}{2} \\ z \end{array}
\right)
$$
$$
\left( \begin{array}{c} A_{s+1} \\ \Phi_{s+\frac{1}{2}} \qquad
\Psi_{s+\frac{1}{2}} \\ B_s \end{array} \right) \quad \Rightarrow
\quad \left( \begin{array}{c} A_{s+1} \\ \Psi_{s+\frac{1}{2}}
\end{array} \right) \quad \oplus \quad \sum_{k=1}^s \quad
\left( \begin{array}{c} \Phi_{k+\frac{1}{2}} \\ A_k \qquad B_k \\
\Psi_{k-\frac{1}{2}} \end{array} \right) \quad \oplus \quad 
\left( \begin{array}{c} \Phi_\frac{1}{2} \\ z \end{array} \right)
$$
It was crucial for the whole construction that each pair of bosonic
fields with equal spins must have opposite parities and one has to
consider a kind of duality mixing between these fields. Moreover such
mixing arises already at the massless supermultiplets level so that
even in the massless infinite spin limit these pairs do not decouple
and we still have two infinite spin bosonic and two infinite spin
fermionic components. It is still possible that by abandoning parity
one can construct the supermultiplet containing just one bosonic and
one fermionic fields but it remains to be checked. 

The mixing angles for the bosonic components take rather different
values for the two type of the supermultiplets but as it can be seen
from their explicit expressions in \cite{Zin07a} in the infinite spin
limit they all become equal. At the same time the main structural
difference between them --- the presence of the left most multiplet
$(A_{s+1},\Phi_{s+1/2})$ --- in the infinite spin limit disappears so
both type of massive supermultiplets produce the same result (up to
some field redefinitions).

\section{Infinite spin fields in $d \ge 5$}

Contrary to the three and four dimensional cases in $d \ge 5$ there
exist an infinite number of such infinite spin representations. Let us
briefly remind how their classification arises \cite{BKRX02}. For the
massless fields we have $p_\mu{}^2 = 0$ and by the Lorentz
transformations one can always bring this vector to the canonical form
$p_\mu = (1,0,\dots,0,1)$. This leads to the so called little group
(i.e. group of transformations leaving this vector intact) that
besides the group $SO(d-2)$ contains pseudo translations $T_i$, $i =
1,2 \dots, d-2$ that are specific combinations of spatial rotations
and Lorentz boosts. Usual finite helicities massless representations
correspond to the case where all $T_i = 0$ while to construct infinite
spin representations one can follow the same root as for the Poincare
group itself. Namely one can consider eighen vectors for this pseudo
translations $T_i |\xi_i> = \xi_i |\xi_i>$, $\xi_i{}^2$ being
invariant. By using $SO(d-2)$ transformations one can always bring
such vector to the form $(1,0,\dots,0)$ and this in turn leads to the
so called short little group $SO(d-3)$ leaving this vector intact.
Thus infinite spin representations are determined by the corresponding
representations of this short little group.

Now it is clear that for the $d=3$ and $d=4$ this short little group
is trivial that is why we have just one bosonic and one fermionic
representations while in $d \ge 5$ there exist infinitely many ones.
For example in $d=5$ and $d=6$ such representations can be labeled by
the parameter $l$ taking integer $l=0,1,2,\dots$ or half integer
$l=\frac{1}{2},\frac{3}{2},\dots$ values for the bosonic and fermionic
cases correspondingly. Lagrangian formulation for such representations
can be obtained form the frame-like gauge invariant formulation for
the massive mixed symmetry bosonic and fermionic fields corresponding
to the Young tableau $Y(k,l)$ with two rows developed in 
\cite{Zin08c,Zin09b,Zin09c}. Namely one has to consider a limit where
mass goes to zero, $k$ goes to infinity while $l$ being fixed. This
construction will be presented in the forthcoming publication so here
let us just illustrate how the spectrum of such representations looks
like (by the spectrum we mean a collection of usual massless fields
that we have to combine to obtain infinite spin one).

Completely symmetric case considered before corresponds to the $l=0$ 
and has the following spectrum (dot stands for the scalar field):
$$
\begin{array}{ccccccc}
\cdot & \yng(1) & \yng(2) & \yng(3) & \yng(4) & \yng(5) & \dots 
\end{array}
$$
For the first non-trivial case $l=1$ we will have two infinite chains
of components:
$$
\begin{array}{cccccc}
\yng(1,1) & \yng(2,1) & \yng(3,1) & \yng(4,1) & \yng(5,1) & \dots \\ 
\\
\yng(1) & \yng(2) & \yng(3) & \yng(4) & \yng(5) & \dots 
\end{array}
$$
The first line begins with the antisymmetric second rank tensor, the
it contains hook and the whole set of long hooks, while in the second
line we again have completely symmetric tensors starting with the
vector fild this time.

Let us give here one more concrete example for $l=3$:
$$
\begin{array}{cccccc}
\yng(3,3) & \yng(4,3) & \yng(5,3) & \yng(6,3) & \yng(7,3) & \dots \\
\\
\yng(3,2) & \yng(4,2) & \yng(5,2) & \yng(6,2) & \yng(7,2) & \dots \\
\\
\yng(3,1) & \yng(4,1) & \yng(5,1) & \yng(6,1) & \yng(7,1) & \dots \\
\\
\yng(3) & \yng(4) & \yng(5) & \yng(6) & \yng(7) & \dots \\
\end{array}
$$
Hopefully the general pattern is clear now. In general in the upper
left corner we have a rectangular diagram with length $l$. Moving to
the right we add one box to the first row, while moving down we cut
one box from the second row until we end again with the completely
symmetric tensors in the bottom line.

\section*{Acknowledgments}

Author is grateful to the I. L. Buchbinder and T. V.
Snegirev for collaboration. Also author is grateful to the
organizers of the "Workshop on higher spin gauge theories", 26-28
April 2017, UMONS, Mons, Belgium for the kind hospitality during the
workshop.

\end{document}